\documentclass[12pt,preprint]{aastex}
\usepackage{natbib}
\usepackage{ifthen}
\newcounter{address}
\newcommand{\latin}[1]{\textit{#1}}
\newcommand{\ie}{\latin{i.e.}}
\newcommand{\eg}{\latin{e.g.}}

\newcommand{\etal}{\latin{et~al.}}
\newcommand{\lcdm}{$\Lambda$CDM}

\begin{document}
\title{
  Cosmic homogeneity demonstrated with luminous red galaxies}
\author{
  David~W.~Hogg\altaffilmark{\ref{NYU},\ref{email}},
  Daniel~J.~Eisenstein\altaffilmark{\ref{Steward}},
  Michael~R.~Blanton\altaffilmark{\ref{NYU}},
  Neta~A.~Bahcall\altaffilmark{\ref{Princeton}},
  J.~Brinkmann\altaffilmark{\ref{APO}},
  James~E.~Gunn\altaffilmark{\ref{Princeton}}, and
  Donald~P.~Schneider\altaffilmark{\ref{PennState}}
}
\setcounter{address}{1}
\altaffiltext{\theaddress}{\stepcounter{address}\label{NYU}
Center for Cosmology and Particle Physics, Department of Physics,
New York University, 4 Washington Pl, New York, NY 10003}
\altaffiltext{\theaddress}{\stepcounter{address}\label{email}
\texttt{david.hogg@nyu.edu}}
\altaffiltext{\theaddress}{\stepcounter{address}\label{Steward}
Steward Observatory, 933 N Cherry Ave, Tucson, AZ 85721}
\altaffiltext{\theaddress}{\stepcounter{address}\label{Princeton}
Princeton University Observatory, Princeton, NJ 08544}
\altaffiltext{\theaddress}{\stepcounter{address}\label{APO}
Apache Point Observatory, P.O. Box 59, Sunspot, NM 88349}
\altaffiltext{\theaddress}{\stepcounter{address}\label{PennState}
Department of Astronomy and Astrophysics,
Pennsylvania State University, University Park, PA 16802}

\begin{abstract}
We test the homogeneity of the Universe at $z\sim 0.3$ with the
Luminous Red Galaxy (LRG) spectroscopic sample of the Sloan Digital
Sky Survey.  First, the mean number $N(R)$ of LRGs within completely
surveyed LRG-centered spheres of comoving radius $R$ is shown to be
proportional to $R^3$ at radii greater than $R\sim
70\,h^{-1}~\mathrm{Mpc}$.  The test has the virtue that it does not
rely on the assumption that the LRG sample has a finite mean density;
its results show, however, that there \emph{is} such a mean density.
Secondly, the survey sky area is divided into 10 disjoint solid
angular regions and the fractional rms density variations of the LRG
sample in the redshift range $0.2<z<0.35$ among these ($\sim
2\times10^7\,h^{-3}~\mathrm{Mpc^3}$) regions is found to be 7~percent
of the mean density.  This variance is consistent with typical biased
\lcdm\ models and puts very strong constraints on the quality of SDSS
photometric calibration.
\end{abstract}


\section{Introduction}

One of the principal assumptions of successful physical cosmological
models is \emph{homogeneity;} \ie, the assumption that sufficiently
large independent volumes of the Universe will contain similar mean
densities of matter (and everything else).  In detail, any test of
this assumption becomes a quantitative one: Do the observed variations
of density agree with the predictions of the leading physical
theories?  Inasmuch as astronomical observations are used to
\emph{rule out} physical theories, the homogeneity of the Universe
cannot be demonstrated definitively beyond this.  This is not because
the observations suggest an inhomogeneous Universe, but rather because
there are no \emph{physical} inhomogeneous models to rule out!

In fact the observations do strongly suggest homogeneity, and the
great success of the \lcdm\ model in explaining the statistical
fluctuations in the isotropic cosmic microwave background \citep[CMB;
\eg,][]{bennett03a}, the growth of present-day large-scale structure
\citep[\eg,][]{tegmark04a}, and the tight redshift-distance relation
inferred for type Ia supernovae \citep{schmidt98a, perlmutter99a}
should be taken as very strong evidence that the Universe is
homogeneous on large scales.  It does not make sense to postulate
inhomogeneous matter distributions without providing a physical model
in which such distributions are consistent with modern observations.

Nevertheless, in what follows we use the enormous Luminous Red Galaxy
(LRG) sample \citep{eisenstein01a} of the Sloan Digital Sky Survey
\citep[SDSS;][]{york00a} to test the homogeneity of the Universe with
the most conservative statistical tests we know.  We may not have
physical inhomogeneous models to test, but such models \emph{may}
exist in the future, and homogeneity on large scales \emph{is} an
extremely strong prediction of \lcdm\ and its variants.  Even if
homogeneity of the matter distribution is taken for granted,
homogeneity in the galaxy distribution is not guaranteed if there are,
\eg, factors contributing to galaxy formation that act over large
distances.  For these reasons, cosmic homogeneity is worthy of study.

Certainly the prevailing view in modern cosmology is that homogeneity
has been very well established by the observed isotropy of the CMB
\citep{partridge67a,wilson67a,smoot92,bennett03a}, x-ray background
\citep[\eg,][]{scharf00a}, and the isotropies of various source
populations, \eg, radio galaxies \citep{peebles93a}.  In the context
of physical models in which the CMB is emitted during recombination at
early times, its isotropy is indeed a very strong argument for
homogeneity.  However it is certainly possible to imagine
inhomogeneous distributions of finite-sized (redshift-dependent)
blackbody sources such that every line of sight happens to terminate
on the surface of one of them.  In other words, isotropy does not by
itself guarantee homogeneity.  The fundamental issue is that
homogeneity is a property of distributions in three-dimensional space
and isotropy is a property of distributions on a two-dimensional sky.

Isotropy of a population of discrete sources merits additional
discussion.  As with the CMB, the isotropy of the point distribution
itself does not guarantee homogeneity, as there can in principle be
inhomogeneous three-dimensional distributions that project to
isotropic distributions on the sky at high probability for typical
observers \citep{durrer97a}.  However, we do expect that isotropy of
such discrete populations provides strong evidence for homogeneity
when combined with the observation that flux and redshift
distributions are similar in different directions.  Isotropy itself is
a two-dimensional test and therefore not sufficient, but isotropy as a
function of flux or redshift is indeed a three-dimensional test and
probably does establish homogeneity, although at lower precision than
the tests we perform below.  Of course these statements (or their
contraries) are hard to maintain with great confidence in the absence
of any inhomogeneous physical model.

By far the cleanest tests of homogeneity involve simply counting
sources in three dimensional regions.  Homogeneity is established---in
principle---when the three-space correlation function of galaxies can
be shown to vanish over a range of scales at large scales.
Unfortunately, most investigations of the correlation function use the
largest scales available in the survey under analysis to determine the
mean density (sometimes with an integral constraint correction).  The
estimation of clustering statistics involves subtraction of the mean
density, so it is in some sense required by the methodology that the
clustering tends to zero at the largest available scales.  Indeed, it
has sometimes been the case that clustering amplitudes have been
underestimated in small surveys.  For these reasons the vanishing of
the correlation function only establishes homogeneity if it is shown
to vanish over a substantial \emph{range} of scales, \eg, as it has
been shown to do for optically selected QSOs on scales of 100 to
$2000\,h^{-1}~\mathrm{Mpc}$ \citep{croom04a}.

The Sloan Digital Sky Survey Luminous Red Galaxy sample is ideal for
performing these tests in an extremely conservative manner.  The
sample used here fills a huge volume ($\sim
0.6\,h^{-3}~\mathrm{Gpc^3}$) with high spectroscopic completeness.  It
contains only galaxies of a limited range of luminosities and it is
close to volume-limited \citep{eisenstein01a}.  The sample contains
large numbers; there are 55,000 used below.  The SDSS footprint also
has a good angular shape for this project; we show below that the LRG
sample contains many independent filled spheres of comoving radius
$R>100\,h^{-1}~\mathrm{Mpc}$.  Previous work on cosmic homogeneity,
some of which has concluded that there is evidence for a fractal-like
galaxy distribution \citep{syloslabini98a,joyce99a}, have not had
samples that have measured such large scales so cleanly.

In what follows, a cosmological world model with
$(\Omega_\mathrm{M},\Omega_\mathrm{\Lambda})=(0.3,0.7)$ is adopted,
and the Hubble constant is parameterized
$H_0=100\,h~\mathrm{km\,s^{-1}\,Mpc^{-1}}$, for the purposes of
calculating distances\citep[\eg,][]{hogg99cosm}.

\section{The LRG sample}

The SDSS \citep{york00a,stoughton02a,abazajian03a,abazajian04a} is
conducting an imaging survey of $10^4$ square degrees in 5 bandpasses
$u$, $g$, $r$, $i$, and $z$ \citep{fukugita96a,gunn98a}.  Photometric
monitoring \citep{hogg01a}, image processing
\citep{lupton01a,stoughton02a,pier03a}, and good photometric
calibration \citep{smith02a} allow one to select galaxies
\citep{strauss02a,eisenstein01a}, quasars \citep{richards02a}, and
stars for follow-up spectroscopy with twin fiber-fed
double-spectrographs.  The spectra cover 3800\AA\ to 9200\AA\ with a
resolution of 1800.  Targets are assigned to plug plates with a tiling
algorithm that ensures nearly complete samples \citep{blanton03a}.

We focus here on the luminous red galaxy spectroscopic sample
\citep{eisenstein01a}.  This uses color-magnitude cuts in $g$, $r$,
and $i$ to select galaxies that are likely to be luminous early-type
galaxies at redshifts between 0.15 and 0.5.  The selection is highly
efficient and the redshift success rate is excellent.  The sample is
constructed so as to be close to volume-limited up to $z=0.36$, with a
dropoff in density towards $z=0.5$.  The comoving number density of
the sample is close to that required to maximize the signal-to-noise
ratio on the large-scale power spectrum.

The sample we use here is drawn from NYU LSS {\tt sample14}
\citep{blanton04a} and covers 3,836 square degrees containing 55,000
LRGs between redshift of 0.16 and 0.47.  We use an absolute $g$-band
magnitude cut of $-21.2$, including $k$-corrections and evolution to
$z=0.3$.  The details of the radial and angular selection functions
are described in \citet{zehavi04a}.  The radial modeling of the
expected number of galaxies as a function of redshift is based closely
on the observed distribution.  The exact survey geometry is expressed
in terms of spherical polygons.  We exclude the few regions that have
less than 48~percent spectroscopy coverage.  We weight all of our
counts by the inverse of the angular selection function, including an
explicit correction for fiber collisions, to bring the weighted LRG
catalog to a uniform mean angular density within the chosen sky
region.

We create large catalogs of randomly distributed points based on these
angular and radial models.  These catalogs match the distribution of
the LRGs in redshift and are isotropic within the survey region.
These catalogs allow us to check the survey completeness of any given
volume and provide a homogeneous baseline (\eg, expected numbers) for
the tests that follow.

The LRG sample depends considerably on the photometric uniformity of
the SDSS.  Photometric calibration is involved at two different stages
in this work: The first stage is at initial target selection, because
the LRG spectroscopic targets are selected on the imaging photometry,
calibrated via the Photometric Telescope \citep{smith02a}. An 0.01 mag
shift in $(r-i)$ color or a 0.03 mag shift in $r$ magnitude would
modulate the LRG target density by up to 10~percent at most redshifts
\citep{eisenstein01a}.  The second stage comes when the spectroscopic
LRG sample is ``cut down'' to the volume limited sample used here by
an additional absolute magnitude cut.  This cut is based on improved
calibration (using survey scan stripe overlap regions).  The sample
density is very sensitive to small shifts in calibration at this stage
also.

The homogeneity that we find below is a testament to the calibration
of the survey.  It appears that while the survey may have small
regions or even scan stripes that are miscalibrated (at $\sim
2$~percent rms), the large-scale photometric homogeneity is better
than 1~percent in $(r-i)$ color, \ie, the errors average down on large
scales.  At no point is the uniformity of the LRG sample assumed or
used in setting calibration or calibration parameters.

\section{Conditioned density scaling}

The fundamental test of the homogeneity of a point set, which makes no
assumption about even the existence of a mean density, is the
measurement of the scaling of the average number of neighboring points
to any given point, as a function of maximum separation
\citep{pietronero87a,gabrielli04a}.  In a homogenous distribution, at
large enough distances, this number $N$ scales with the maximum
distance $R$ as $R^D$, where $D$ is the dimensionality of the space.

\begin{figure}
\plotone{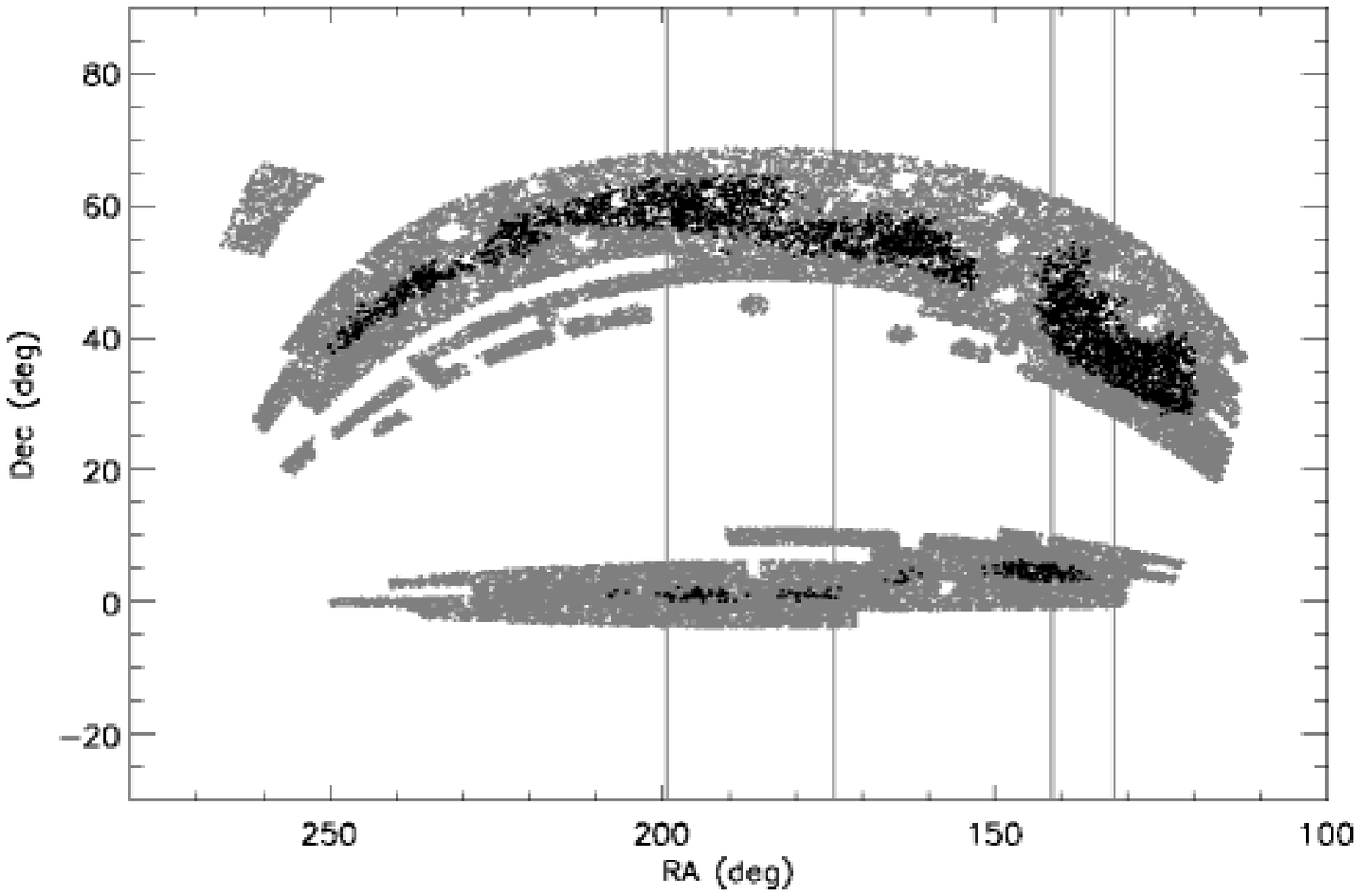}
\caption{The entire LRG sample used in this study is shown in grey.
The 3658 LRGs in the redshift range $0.2<z<0.4$ with $>95$~percent
complete volumetric coverage inside the surrounding comoving spheres
of radii $R=50$ and $100\,h^{-1}~\mathrm{Mpc}$ are shown in black.
Also shown are grey lines that split the 3658 points into five
quantiles by RA.\label{fig:sky}}
\end{figure}

Because our goal is to be extremely conservative, for this test, we
considered neighbors not of all LRGs, but only of those ``target''
LRGs in the redshift range $0.20<z<0.40$ that have enormous spherical
comoving regions around them that have been completely surveyed by the
SDSS spectroscopic survey.  In detail, we required that \emph{both}
the $50$ \emph{and} the $100\,h^{-1}$~Mpc radius comoving sphere
around each target galaxy to be at least 95~percent complete (in the
sense of 95~percent coverage by the spectroscopic survey, which itself
is complete at the 94~percent level for LRGs).  The sky distribution
of the 3658 LRGs that satisfy this highly restrictive selection are
shown in Figure~\ref{fig:sky}.

\begin{figure}
\plotone{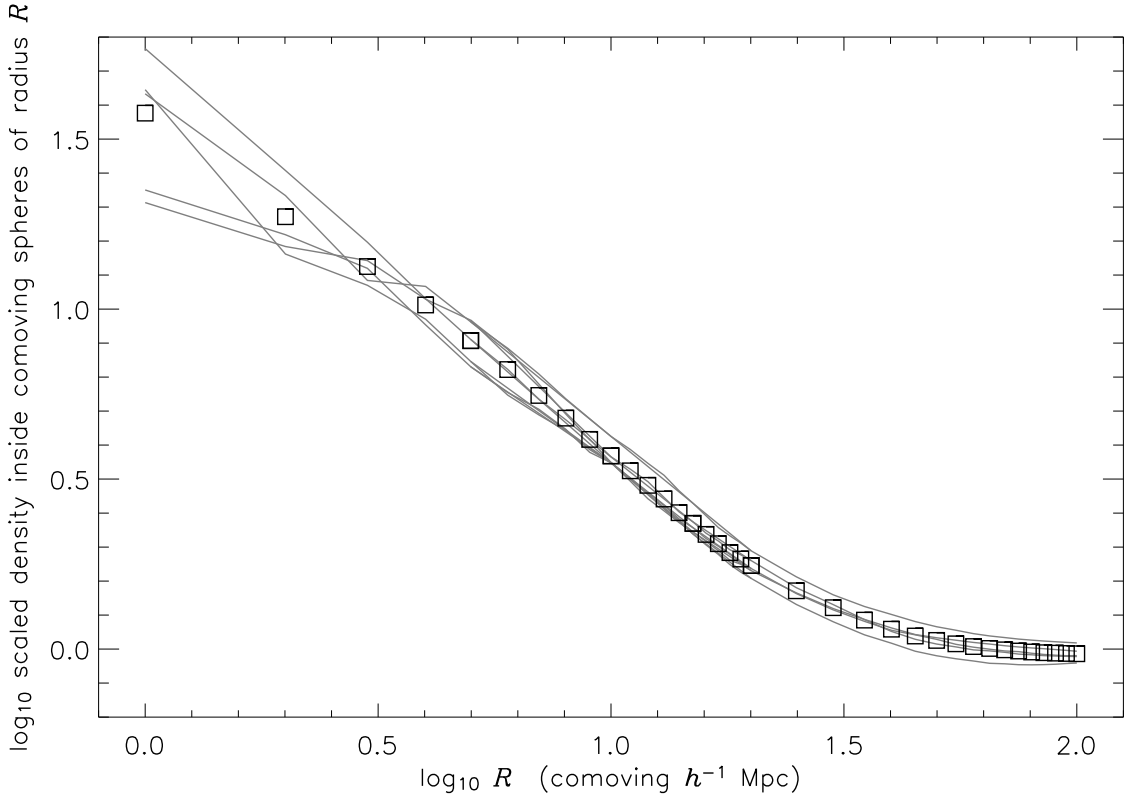}
\caption{The average comoving number density (\ie, number counted
divided by expected number from a homogeneous random catalog) of LRGs
inside comoving spheres centered on the 3658 LRGs shown in
Figure~\ref{fig:sky}, as a function of comoving sphere radius $R$.
The average over all 3658 spheres is shown in with open squares, and
the averages of each of the five RA quantiles is shown as a separate
grey line.  At small scales, the number density drops with radius
because the LRGs are clustered; at large scales the number density
approaches a constant because the sample is
homogeneous.\label{fig:density}}
\end{figure}

In Figure~\ref{fig:density} the scaling of the average number $N$ of
neighbors of the target galaxies as a function of radius $R$ is shown,
with an ``expected number'' divided out.  The expected number is
estimated by counting the number of ``random'' points in the spherical
volume, where the random points are from a homogeneous random catalog
described above with the same footprint, angular variation of
completeness, and radial selection function as the LRGs, but 100 times
the total number.  Since the sample being used is highly complete,
this is essentially equivalent to dividing all the counts by a
constant times $R^3$.  Figure~\ref{fig:density} shows that the mean
number $N$ scales as $R^3$ at large distances (and $R^2$ at small
distances).

Also shown in Figure~\ref{fig:density} is the average number for each
of five samples, split by RA by the RA lines shown in
Figure~\ref{fig:sky}.  Each of the five sky regions shows the $R^3$
scaling individually.  The expected number densities from the random
catalogs has been kept constant across all regions.  This test is
particularly conservative, because the five sky regions contain
(mostly) independent spheres and are expected to have different
observing and calibration properties.

\section{Density variation}

In an inhomogeneous universe (absent a physical inhomogeneous model),
the natural expectation is for order unity differences in population
densities on all scales.  For this reason we performed a test in which
the sky is split into disjoint regions and the number density of LRGs
is tested in a comoving volume in each of the disjoint regions.

\begin{figure}
\plotone{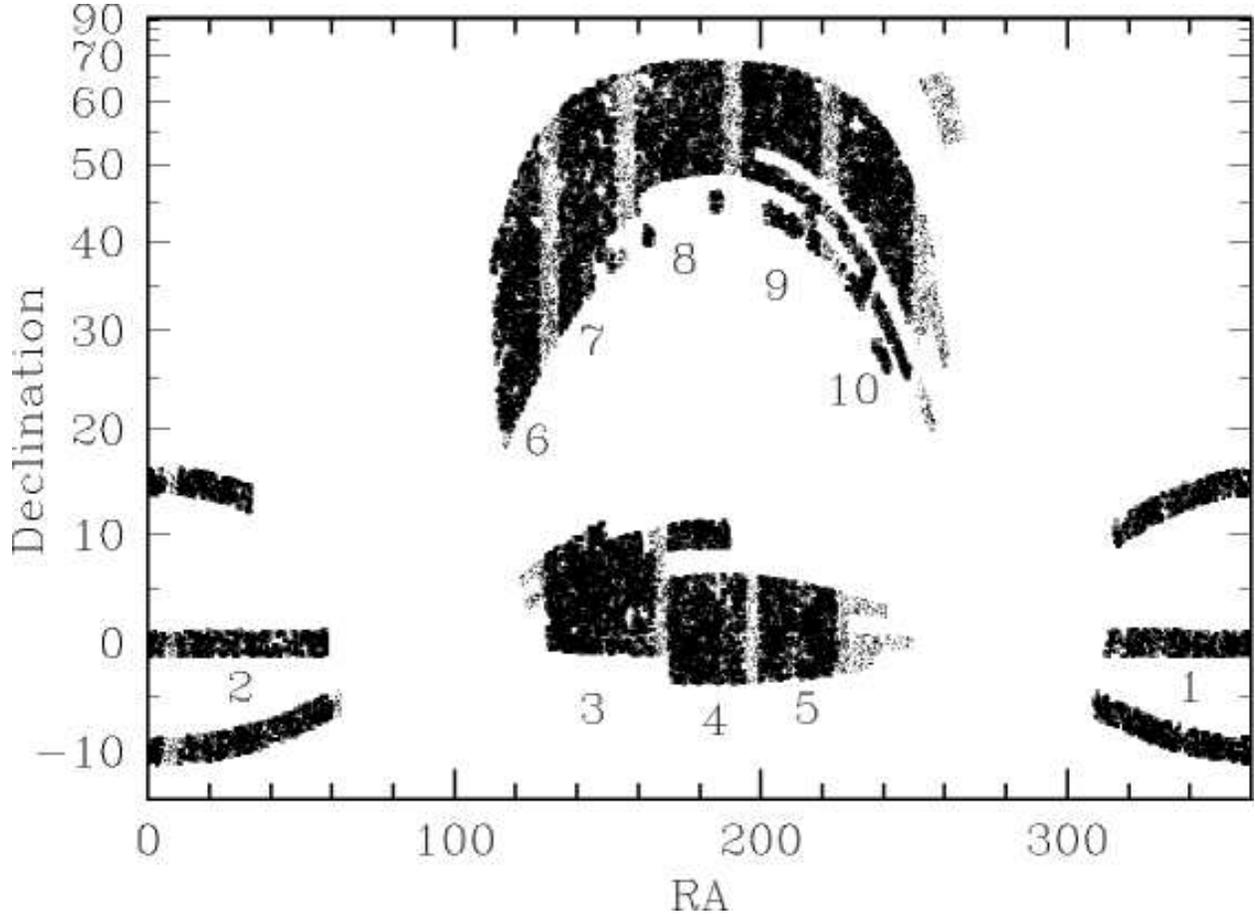}
\caption{The LRGs in the redshift range $0.20<z<0.35$, divided into 10
disjoint sky regions (where LRGs are labeled and marked by dots and
squares), with inter-region gaps (where LRGs are marked by dots
only).\label{fig:squeeze}}
\end{figure}

\begin{figure}
\plotone{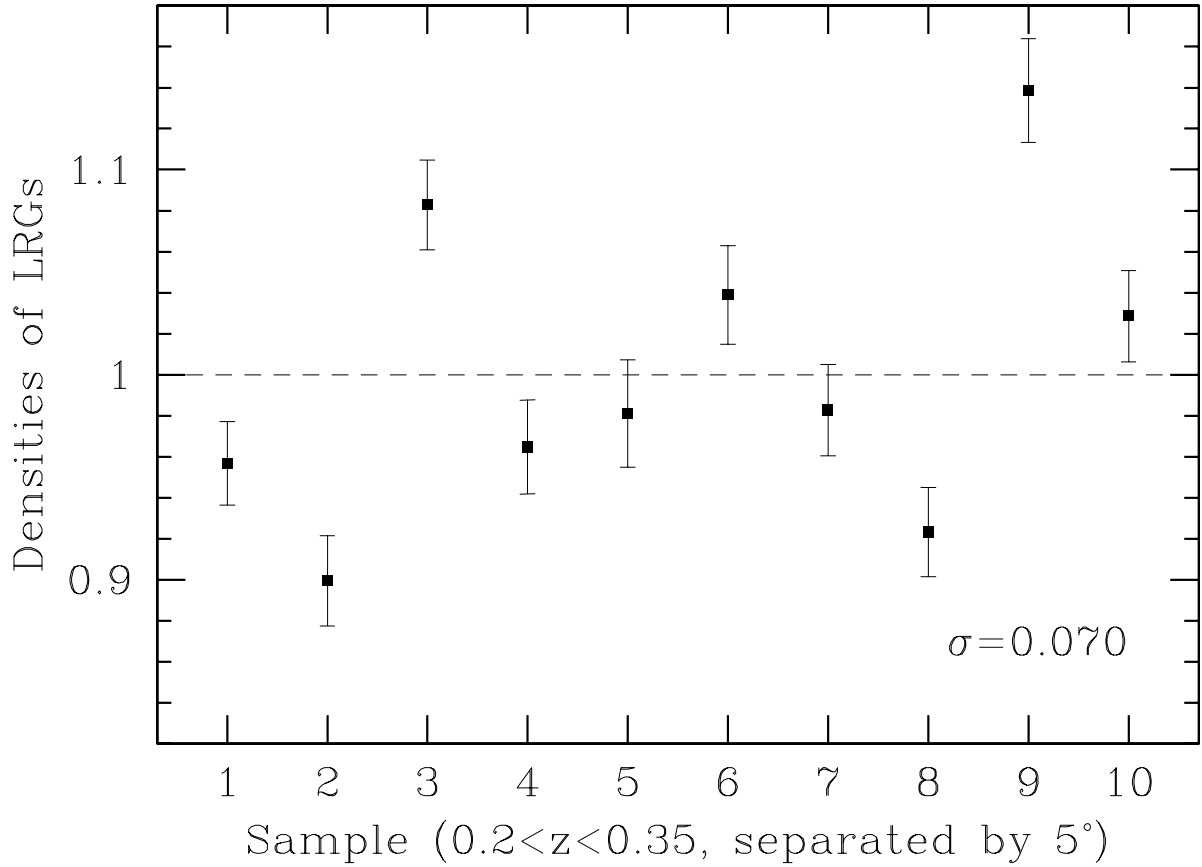}
\caption{The relative comoving number densities of LRGs in the
redshift range $0.20<z<0.35$ in the 10 disjoint sky areas shown in
Figure~\ref{fig:squeeze}.  The quoted ``$\sigma$'' is the rms scatter
of the points, with the expected Poisson rms subtracted out in
quadrature.\label{fig:counts}}
\end{figure}

Figure~\ref{fig:squeeze} shows the redshift $0.20<z<0.35$ LRG sample
split into 10 disjoint sky regions, each of which is of comparable
total solid angle and therefore (for this redshift range) comoving
volume.  Each region corresponds to a comoving volume of roughly
$2.2\times 10^7\,h^{-3}~\mathrm{Mpc^3}$.  Figure~\ref{fig:counts}
shows the relative densities of these regions.  The rms scatter
between the 10 regions is 7.3~percent.  Subtracting the expected
Poisson variation among these regions yields a relative rms scatter of
7.0~percent.

In a $\Lambda$CDM model with $\sigma_8=0.85$, the mass density
fluctuations among this set of sky regions in the redshift range
$0.20<z<0.35$ is expected to be 3.1~percent.  The LRG real-space
autocorrelation has been measured to have an amplitude $\sigma_8=1.80$
\citep{zehavi04a}, so that the galaxies have a bias of around 2, and
one would predict a variation of roughly 6.5~percent.  Hence, the
7.0~percent variance in the LRG sample does not leave much room for
calibration errors, which ought to come in in quadrature.  For the LRG
sample, 1~percent calibration variations in the $r-i$ color or
3~percent variations in the $r$ flux would produce $\sim 10$~percent
variations in the LRG number density \citep{eisenstein01a}, so the
consistency of this measurement with the biased CDM description of the
LRG population puts extremely strong constraints on the quality of
SDSS photometric calibration.  Note that the LRG target selection is
never used to tune the photometric calibration, so there is no sense
in which this uniformity is enforced directly.

\section{Discussion}

The extremely large (in both volume and number) and complete LRG
sample from the SDSS was used to test the three-dimensional
homogeneity (\ie, not just isotropy) of the Universe at $z\sim 0.3$ by
the most conservative possible method.  We find that the Universe has
a well-defined mean density and that it is homogeneous.  Furthermore,
the variations we see in the density of LRGs on large scales is
consistent with the predictions of a biased \lcdm\ cosmogonic model.

The number $N(R)$ of LRGs within LRG-centered spheres of comoving
radius $R$ is shown to be proportional to $R^3$ at radii greater than
$R\sim 70\,h^{-1}~\mathrm{Mpc}$.  For this test we used not all LRGs
as central LRGs, but only those in the centers of highly complete
spheres, so there is no ``interpolation'' or ``extrapolation'' over
unobserved regions.  In the terminology of fractals, this test shows
that the fractal dimension of the LRG distribution is very close to
$D=3$.  The test has the virtue that it does not rely on the
assumption that the LRG sample has a finite mean density; its results
show, however, that there \emph{is} such a mean density.

The $D=3$ result presented here is in qualitative disagreement with
some previous studies.  There are some similar kinds of analyses of
redshift surveys that show $D\approx 2$ \citep{syloslabini98a,
joyce99a}, but there is no quantitative disagreement, because the most
robust measurements of $D$ are generally made at scales much smaller
than those measured by the LRGs here.  Indeed, Figure~2 shows that
$D=2$ is a remarkably good fit out to roughly
$20\,h^{-1}~\mathrm{Mpc}$.  Analyses of the ESO Slice Project, in
which the scaling of the number of galaxies in the sample is measured
as a function of the redshift depth to which they are counted
\citep{scaramella98a,joyce99b}, depend crucially on $K$ corrections,
evolution, and world model.

The survey sky area was divided into 10 disjoint solid angular regions
and the fractional rms density variations of the LRG sample in the
redshift range $0.2<z<0.35$ among these ($\sim
2\times10^7\,h^{-3}~\mathrm{Mpc^3}$) regions was found to be 7~percent
of the mean density.  This variance is consistent with typical biased
\lcdm\ models, with a bias of $\approx 2$, as is found for the LRG
sample at smaller scales \citep{zehavi04a}.  This result confirms
homogeneity and supports the biased \lcdm\ model on the largest
observable scales.

Finally, it is worthy of note that because the LRG sample selection is
so sensitive to photometric calibration, these results demonstrate
that the calibration of the SDSS in the $g$, $r$, and $i$ bands is
consistent at the sub-percent level when averaged on angular scales of
tens of degrees and larger.  This is a tremendous technical
achievement and recommends the LRG sample and the SDSS for making many
extremely precise measurements in the future.

\acknowledgments It is a pleasure to thank Michael Joyce and Francesco
Sylos Labini for the discussions which led to this project and the
Laboratoire de Physique Th\'eorique, Universit\'e de Paris XI, for
hospitality during that time.  Additional thanks are due to Jim
Peebles and Michael Strauss for useful discussions.  DWH and MRB are
partially supported by NASA (grant NAG5-11669) and NSF (grant
PHY-0101738).  DJE is supported by NSF (grant AST-0098577) and an
Alfred P.~Sloan Research Fellowship.

Funding for the creation and distribution of the SDSS Archive has been
provided by the Alfred P. Sloan Foundation, the Participating
Institutions, the National Aeronautics and Space Administration, the
National Science Foundation, the U.S. Department of Energy, the
Japanese Monbukagakusho, and the Max Planck Society. The SDSS Web site
is http://www.sdss.org/.

The SDSS is managed by the Astrophysical Research Consortium (ARC) for
the Participating Institutions. The Participating Institutions are the
University of Chicago, Fermilab, the Institute for Advanced Study, the
Japan Participation Group, the Johns Hopkins University, the Korean
Scientist Group, Los Alamos National Laboratory, the
Max-Planck-Institute for Astronomy (MPIA), the Max-Planck-Institute
for Astrophysics (MPA), New Mexico State University, the University of
Pittsburgh, Princeton University, the United States Naval Observatory,
and the University of Washington.

\clearpage

\clearpage

\clearpage

\clearpage

\end{document}